\documentclass[twocolumn,groupedaddress,nofootinbib]{revtex4}
\usepackage{graphicx,amsfonts,amsmath,color,epsfig,epstopdf}
\newcommand{\SU}[1]{\ensuremath{\mathrm{SU}( #1 )}}
\newcommand{\Un}[1]{\ensuremath{\mathrm{U}( #1 )}}
\newcommand{\SO}[1]{\ensuremath{\mathrm{SO}( #1 )}}

\newcommand{\SpR}[1]{\ensuremath{\mathrm{Sp}( #1,\mathbb{R} )}}
\newcommand{\CG}[3]{\ensuremath{\langle#1;#2|\,#3\rangle}}

\newcommand{\ket}[1]{\ensuremath{\left| #1 \right\rangle}}

\newcommand{\braketop}[3]{\ensuremath{\left\langle #1 | #2 | #3 \right\rangle}}

\newcommand{\RedME}[3]{\ensuremath{\langle #1 \| #2 \| #3 \rangle}}
\newcommand{\RedMEnonNorm}[3]{\ensuremath{( #1 \| #2 \| #3 )}}

\newcommand{\half}{\ensuremath{\textstyle{\frac{1}{2}}}}

\newcommand{\betb}{\begin{tabular}{p{4.0cm}p{9.0cm}}}
\newcommand{\entb}{\end{tabular}}

\newcommand{\ho}{\ensuremath{\hbar\Omega}}
\newcommand{\ph}[1]{\ensuremath{#1}p-\ensuremath{#1}h}
\begin{document}

\title{
Symplectic No-core Shell-model Approach to Intermediate-mass Nuclei
}
\author{G. K. Tobin}
\author{M. C. Ferriss}
\author{K. D. Launey}
\author{T. Dytrych}
\author{J. P. Draayer}
\author{A. C. Dreyfuss}
\affiliation{Department of Physics and Astronomy, Louisiana State
University, Baton
Rouge, LA 70803, USA}
\author{C. Bahri}
\affiliation{Department of Physics, University of Notre Dame,
Notre Dame, Indiana 46556-5670, USA}

\begin{abstract}
We present a microscopic description of nuclei in an  intermediate-mass region, including the proximity to the proton drip line, based on a no-core shell model with a schematic many-nucleon long-range interaction with no parameter adjustments. The outcome confirms the essential role played by the symplectic symmetry to inform the interaction and the winnowing of shell-model spaces. We show that it is imperative that model spaces be expanded well beyond the current limits up through fifteen major shells to accommodate particle excitations that appear critical to highly-deformed spatial structures and the convergence of associated observables.

\end{abstract}


\maketitle

\section{Introduction}

For intermediate-mass (`$sd$-shell' ) nuclei, which are currently inaccessible 
by standard {\it ab initio} no-core shell-model   (NCSM) \cite{NCSM} calculations, symmetry-based considerations  are essential.  In particular, we employ the no-core symplectic (NCSpM) shell model for symmetry-preserving interactions  \cite{DreyfussLDDB13} with \SpR{3} the underpinning symmetry \cite{RosensteelR77}. This symmetry is inherent to the dynamics of deformed nuclear systems \cite{RosensteelR80,DraayerWR84,Rowe85,BahriR00,DytrychSBDV06}. The present study uses a schematic, but fully microscopic and physically motivated effective many-nucleon interaction, a choice that enables the use of group-theoretical methods with analytical expressions for Hamiltonian matrix elements, and which in turn makes large space solutions for $sd$-shell nuclei feasible. 

Recently, we successfully applied the NCSpM model to the rotational and alpha-cluster substructures of $^{12}$C, including the Hoyle state (the second $0^+$ state in $^{12}$C) and its rotational band  \cite{DreyfussLDDB13}, as well as of $^{8}$Be \cite{LauneyDDTFLDMVB12}. The symplectic model has been used previously to achieve a remarkable reproduction of enhanced $E2$ transition strengths in $^{20}$Ne without effective charges and with the use of a relatively simple symmetry-breaking valence-shell interaction \cite{DraayerWR84}.  In addition, it has been applied to  $^{166}$Er using the Davidson potential \cite{BahriR00}.  

The main objective of the present study  is to offer qualitative results that can provide  guidance for {\it ab initio} shell model  approaches by informing key features of nuclear structure and the interaction, first on 
the physically relevant truncation of shell-model spaces, but also on the dominant deformation and particle-hole configurations. This is especially useful for the {\it ab initio} symmetry-adapted  no-core shell model (SA-NCSM) \cite{DytrychPANP2013}, which will then bring forward, with the use of a realistic nucleon-nucleon interaction,  an  accurate reproduction and reliable prediction of energy spectra and associated transition rates that majorly impact modeling of stellar explosions and astrophysical processes.

In this study, we explore the ground-state ($g.st.$) rotational band of  lower $sd$-shell nuclei, namely, $^{20}$O, $^{20,22,24}$Ne, $^{20,22}$Mg, and $^{24}$Si. These low-lying states are expected to be highly influenced by large deformation. This, together with the combinatorial growth in model space dimensionality with number of particles and the spaces in which they primarily reside, has hitherto precluded a no-core shell model description and typically, in this region, valence shell model or mean field  approaches have been employed (e.g., \cite{Brown01,CaurierMP05,Hinohara11}). Many of these nuclei are in close proximity to the proton drip line and are key to understanding, e.g., novae and X-ray bursts (see, e.g., \cite{BlackmonAS06}). In particular, properties of low-lying $2^+$ and $4^+$ states in isotopes as $^{20}$Mg and $^{24}$Si are required to predict $(p,\gamma)$ reaction rates that are expected to affect the light curve for X-ray bursts. As such unstable isotopes are very hard to make experimentally and state-of-the-art radioactive-beam measurements have only recently started to provide new information  \cite{BlackmonAS06}, theoretical predictions are valuable.

The present approach utilizes symmetry to reduce the  dimensionality of the model space through a very structured winnowing of the 
basis states to physically relevant subspaces. 
Indeed, experimental evidence supports the fact that in this mass range, the dynamics 
favors a dominance of  low spin and high deformation, which has been demonstrated by symmetry-guided theoretical studies \cite{RosensteelR80,DraayerWR84,Rowe85} as well as through an { \it ab initio}
study \cite{DytrychSBDV06,DytrychDSBV09}. The latter exploits symplectic symmetry and its deformation-related \SU{3} subgroup in an analysis of { \it ab initio} large-scale nuclear physics applications for $^{12}$C and $^{16}$O.  The outcome of this study has revealed that typically only one or two symplectic many-body basis states  (vertical cones) suffice to represent a large fraction -- typically  in excess of about 80\% of the physics -- as measured by projecting { \it ab initio} NCSM results onto a symmetry-adapted equivalent basis. Such a symplectic pattern has been also observed in an { \it ab initio} SA-NCSM study of $^{6}$Li,  $^{6}$He, and $^{8}$Be \cite{DytrychPANP2013}. These findings point to the relevance of the symplectic symmetry, first to the many-body nuclear wavefunctions, and then to the inter-nucleon interaction (as symplectic basis states appear not to mix strongly). The NCSpM builds upon these considerations, and here we offer solutions to $sd$-shell nuclei in  the framework of a fully microscopic no-core shell model. This, in turn, allows us to examine the role of currently inaccessible shell-model spaces, up through 15 major shells, and of associated particle excitations to these shells for a description of large deformation.

\vspace{-0.2in}
\section{Symmetry-informed Approach}
We employ the no-core symplectic model (NCSpM), outlined in Refs. \cite{BahriR00}, with a novel interaction that is effectively realized by  an exponential dependence on the quadrupole-quadrupole ($Q.Q$) two-body interaction,  the physically relevant interaction of each particle with the total quadrupole moment of the nuclear system. This introduces simple but important many-body interactions that enter in a prescribed hierarchical way given in powers of a small parameter, the only adjustable parameter in the model.  The model offers a microscopic no-core shell-model description of nuclei in terms of mixed deformed configurations and allows the inclusion of higher-energy particle excitations \cite{DreyfussLDDB13} that are currently inaccessible by {\it ab initio} shell models.
It reduces to the successful Elliott model \cite{Elliott58ElliottH62} in the limit of a single valence shell and a zero model parameter.  

The underlying symmetry of the NCSpM is the symplectic \SpR{3} group \cite{RosensteelR77} and its embedded \SU{3} subgroup  \cite{Elliott58ElliottH62}. The symplectic basis (detailed in \cite{Rowe85}) utilized in  NCSpM is related,  via a unitary transformation, to the three-dimensional HO ($m$-scheme) many-body basis used in the NCSM  (see the review \cite{DytrychSDBV08}). The conventional NCSM  basis spaces \cite{NCSM} are constructed using HO single-particle states and are characterized by the $\hbar\Omega$ oscillator strength  (or equivalently, the oscillator length $b=\sqrt{ \frac{\hbar }{m\Omega} }$ for a nucleon mass $m$)
as well as  by the cutoff in total oscillator quanta, $N_{\max} $, above the lowest HO energy configuration for a given nucleus. Indeed, the NCSpM employed within a complete model space up through $N_{\max}$, will coincide with the NCSM for the same $N_{\max}$ cutoff. 

The important feature of the NCSpM model is its ability to down-select to the most relevant configurations, which are chosen among all possible \SpR{3} irreducible representations (irreps) within an $N_{\max}$ model space.  The \SpR{3} irreps divide the space into `vertical cones' that are comprised of basis states of definite $(\lambda\,\mu)$ quantum numbers of \SU{3} linked to the intrinsic quadrupole deformation \cite{RosensteelR77b,LeschberD87,CastanosDL88}. E.g., the simplest cases, $(0\, 0)$, $(\lambda\, 0)$, and $(0\,\mu)$,  describe spherical, prolate, and oblate deformation, respectively, while a general nuclear state is typically a superposition of several hundred various $(\lambda\,\mu)$ triaxial deformation configurations.

\vspace{-0.2in}
\subsection{Symplectic \SpR{3} group}
The translationally invariant (intrinsic) symplectic generators  can be written as \SU{3} tensor operators in terms of the harmonic oscillator raising, $b_{i \alpha}^{\dagger(1\,0)}=\frac{1}{\sqrt{2}}(X_{i \alpha}-iP_{i \alpha})$, and lowering $b^{(0\,1)}$ dimensionless operators (with $\mathbf X$ and $\mathbf P$ the lab-frame position and momentum coordinates and $\alpha=1,2,3$ for the three spatial directions), 
\begin{eqnarray}
A^{(2\,0)}_{\mathfrak{L}M}\!\!&=&\!\! 
\frac{1}{\sqrt{2}} \sum_{i=1}^A
\left[b_{i}^{\dagger}\times b_{i}^{\dagger}\right]^{(2\,0)}_{\mathfrak{L}M}
- \frac{1}{\sqrt{2}A} \sum_{s,t=1}^A
\left[b^{\dagger}_{s}\times b^{\dagger}_{t}\right]^{(2\,0)}_{\mathfrak{L}M}
 \label{sp3RgenA}\\
C^{(1\,1)}_{\mathfrak{L}M}\!\!&=&\!\!
\sqrt{2} \sum_{i=1}^A
\left[b_{i}^{\dagger}\times b_{i}\right]^{(1\,1)}_{\mathfrak{L}M}
\!- \frac{\sqrt{2}}{A} \sum_{s,t=1}^A
\left[b^{\dagger}_{s}\times b_{t}\right]^{(1\,1)}_{\mathfrak{L}M},
\label{sp3RgenC}
\end{eqnarray}
together with $B^{(0\,2)}_{\mathcal{L}M}=(-)^{\mathcal{L}-M}(A^{(2\,0)}_{\mathcal{L}-M})^{\dagger}$ ($\mathcal{L}=0,2$) and
$H_{00}^{(00)}= \sqrt{3} \sum_{i} \left[b_{i}^{\dagger}\times b_{i}\right]^{(00)}_{00}
-\frac{\sqrt{3}}{A} \sum_{s,t}
\left[b^{\dagger}_{s}\times b_{t}\right]^{(0\,0)}_{00}
+\frac{3}{2}(A-1)$, 
where the sums run over all $A$ particles of the system. The eight operators $C^{(1\,1)}_{\mathcal{L},M}$ ($\mathcal{L}=1,2$) generate the \SU{3} subgroup of
\SpR{3}. They realize the angular momentum operator:
\begin{equation}
L_{1M}=C^{(1\,1)}_{1M},\, M=0,\pm1,
\label{Lgen}
\end{equation}
 and the Elliott algebraic quadrupole moment tensor $\mathcal{Q}^{a}_{2M}=\sqrt{3}C^{(1\,1)}_{2M},\, M=0,\pm1,\pm2$. The mass quadrupole moment can be constructed in terms of the symplectic generators as,
\begin{equation}
Q_{2M}=\sqrt{3}(A^{(2\,0)}_{2M}+C^{(1\,1)}_{2M}+B^{(0\,2)}_{2M}).
\label{Qgen}
\end{equation}

\vspace{-0.2in}
\subsection{Symplectic basis}
A many-body basis state of a symplectic irrep is labeled according to the group chain,
\begin{equation}
\begin{array}{cccccccc}
\SpR{3}       & \supset & U(3)  & \supset & \SO{3} & \supset & SO(2)     \\
\sigma         & n\rho     & \omega & \kappa  & L    & & M
\end{array}
\end{equation}
and constructed by acting with symmetrically
coupled polynomials in the symplectic raising operators, $A^{(2\,0)}$, on a unique
symplectic bandhead configuration, $\ket{\sigma}$,
\begin{equation}
|\sigma n\rho\omega \kappa L M\rangle 
= \left[ \left[A^{(2\,0)}\times A^{(2\,0)} \dots \times A^{(2\,0)}\right]^{n}\times\ket{\sigma}\right]^{\rho\omega}_{\kappa L M},
\label{basis}
\end{equation}
where $\sigma$ $\equiv $ $N_\sigma\left(\lambda_{\sigma}\, \mu_{\sigma}\right)$
labels the \SpR{3} irrep, $n\equiv N_{n}\left(\lambda_{n}\,
\mu_{n}\right)$, $\omega\equiv N_\omega\left(\lambda_{\omega}\,
\mu_{\omega}\right)$, and  $N_{\omega}=N_{\sigma}+N_{n}$ is
the total number of HO quanta  ($\rho$ and $\kappa$ are multiplicity labels). This can be generalized to include spin,
$|\sigma n\rho\omega \kappa (L S_{\sigma}) JM_J \rangle = \sum_{MM_S} \CG{L M}{S_{\sigma} M_S}{J M_J} |\sigma n\rho\omega \kappa L M S_{\sigma} M_S  \rangle$, and also isospin.
States within a symplectic irrep have the same spin (isospin) value, which is given by the spin $S_\sigma$ (isospin $T_\sigma$) of the bandhead  $|\sigma; S_{\sigma}\rangle$ \cite{DytrychSDBV08}. 
Symplectic basis states span the entire shell-mode space\footnote{A complete set of labels includes additional quantum numbers $\ket{\left\{\alpha \right\}\sigma}$ that distinguish different bandheads with the same $N_\sigma\left(\lambda_{\sigma}\, \mu_{\sigma}\right)$.  \SpR{3}-preserving Hamiltonians render energy spectra degenerate with respect to $\left\{\alpha \right\}$. However, for all present calculations for $g.st.$ rotational bands and associated observables, $\left\{\alpha \right\}$ is unique (an only set).}.

The symplectic  structure accommodates relevant particle-hole (p-h) configurations in a natural way (see also Fig. 1 of Ref. \cite{DreyfussLDDB13}). According to Eq. (\ref{basis}), the basis states of an \SpR{3} irrep (vertical cone) are built over a bandhead $\ket{\sigma}$  by 2\ho~ \ph{1} (one particle raised by two shells)
monopole ($\mathfrak{L}=0$) or quadrupole ($\mathfrak{L}=2$) excitations, realized by the first term in $A^{(2\,0)}_{\mathfrak{L}M}$ of Eq. (\ref{sp3RgenA}),  together with a smaller 2\ho~\ph{2} correction for eliminating the spurious center-of-mass (CM) motion, realized by the second term in $A^{(2\,0)}_{\mathfrak{L}M}$. The symplectic 
bandhead $\ket{\sigma}$ is  the lowest-weight \SpR{3} state, which is defined by the usual requirement that the symplectic lowering operators annihilate it -- in analogy to a $|J,M_J=-J\rangle$ state for the case of the \SU{2} group of  angular momentum, that is, $J_-|J,M_J=-J\rangle=0$.  The  bandhead, $\ket{\sigma; \kappa_\sigma L_\sigma M_\sigma}$, is an \SU{3}-coupled many-body state with a given nucleon distribution over the HO shells and while not utilized here, can be obtained in terms of the creation operators $a^\dagger_{(\eta\, 0)}=a^\dagger_\eta$, which create a particle in the HO shell $\eta=0,1,2,\dots$. E.g., for a 0\ho~bandhead, the nucleon distribution is a single configuration,
\begin{equation}
\left[ a^\dagger_{(\eta_1\, 0)} \times a^\dagger_{(\eta_2\, 0)} \times  \dots \times  a^\dagger_{(\eta_A\, 0)} \right]^{(\lambda_{\sigma}\, \mu_{\sigma})}_{\kappa_\sigma L_\sigma M_\sigma}\ket{0}
\end{equation}
with $N_ \sigma = \eta_1+ \eta_2 + \dots + \eta_A+\frac{3}{2}(A-1)$, such that $N_\sigma \ho$ includes the HO zero-point energy and $3/2$ is subtracted to ensure a proper treatment of the CM. 
To eliminate the spurious CM motion, the NCSpM also uses  symplectic generators constructed in ${\mathbf r}_{i}$ (${\mathbf p}_{i}$) particle position (momentum) coordinates relative to the  CM.
 These generators are used to build the basis, the interaction, the many-particle kinetic energy operator, as well as to evaluate observables. 

An example for the symplectic basis states follows for $^{24}$Mg. Its lowest HO-energy configuration is given by $N_\sigma= 62.5$ or 0\ho, while the 4\ho~$(20\,0)$ symplectic irrep includes:
\begin{enumerate}
\item A  bandhead  ($N_n=0$) with  $N_\sigma= 66.5$ (or 4\ho) and $(\lambda_ \sigma \,\mu_ \sigma)=(20\,0)$;
\item  $N_n=2$  states with $N_ \omega $=68.5 and $(\lambda_ \omega \,\mu_ \omega)= (22\, 0)$, $(20\, 1)$, and $(18\, 2)$;
\item and so forth for higher $N_n$.
\end{enumerate}
For each $(\lambda_ \omega \,\mu_ \omega)$, the quantum numbers $\kappa$, $L$ and $M$ are given by Elliott \cite{Elliott58ElliottH62}. E.g., for $(22\,0)$, $\kappa=0$, $L=0,2,4,\dots,22$, and $M=-L,-L+1,\dots,L$.

\subsection{Symmetry-preserving interactions}
We note that the NCSpM, as presented here, is limited to interactions  that preserve the \SpR{3} symmetry. This restriction facilitates the use of a group-theoretical apparatus and analytical expressions for the Hamiltonian matrix elements, which, in turn, makes it possible to incorporate large $N_{\max}$ spaces in applications of the theory. 
{\it Ab initio} calculations lie beyond the scope of the current analysis, but
the addition of symmetry-mixing terms in the interaction is feasible, and a logical extension of the theory to include such terms is under development.
\SpR{3}-symmetric Hamiltonians appear to be particularly suitable to capture the essential characteristics of the low-energy nuclear kinematics and dynamics.
The reason is that important pieces of the inter-nucleon Hamiltonian of a quantum  many-body system can be expressed in terms of the \SpR{3} generators, which directly relate to the relative particle momentum and position coordinates, as well as straightforwardly account for the Pauli exclusion principle. Indeed, the many-particle kinetic energy ($\sum_i\frac{{\mathbf p}_{i}^2}{2m}$), the HO potential ($\sum_i\frac{m\Omega^2 {\mathbf r}_{i}^2}{2}$), the mass quadrupole moment operator ($Q$), and the orbital momentum ($L$) are all elements of the $\SpR{3}\supset \Un{1}\times \SU{3} \supset\SO{3}$ structure. 

Hence, \SpR{3}-preserving Hamiltonians can include the many-particle kinetic energy:
\begin{equation}
\frac{T}{\ho}=\frac{1}{\ho}\sum_i\frac{{\mathbf p}_{i}^2}{2m}=\frac{1}{2}H_{00}^{(00)} -\sqrt{\frac{3}{8}}(A^{(2\,0)}_{00}+B^{(0\,2)}_{00}),
\end{equation}
the HO potential:
\begin{equation}
\frac{V_{HO}}{\ho}=\frac{1}{\ho}\sum_i\frac{m\Omega^2 {\mathbf r}_{i}^2}{2} =\frac{1}{2}H_{00}^{(00)} +\sqrt{\frac{3}{8}}(A^{(2\,0)}_{00}+B^{(0\,2)}_{00}),
\end{equation}
 as well as terms dependent on  $L$, see Eq. (\ref{Lgen}), and $Q$, see Eq. (\ref{Qgen}). These interactions have analytical matrix elements in the \SpR{3} basis  (\ref{basis}) and act within a symplectic vertical cone ($\sigma_f= \sigma_i \equiv \sigma $). For example, for the dimensionless many-particle kinetic energy, 
$\frac{T}{\ho}$, the matrix elements are given as:
\begin{widetext}
\begin{eqnarray}
&&\braketop{\sigma n_f\rho_f\omega_f \kappa_f L_f M_f}{\frac{T}{\ho}}{\sigma n_i\rho_i\omega_i \kappa_i L_i M_i} \nonumber \\
&&=\frac{1}{2}\braketop{\sigma n_f\rho_f\omega_f \kappa_f L_f M_f}{H_{00}^{(00)}}{\sigma n_i\rho_i\omega_i \kappa_i L_i M_i} 
-\sqrt{\frac{3}{8}}\braketop{\sigma n_f\rho_f\omega_f \kappa_f L_f M_f}{A^{(2\,0)}_{00}+B^{(0\,2)}_{00}}{\sigma n_i\rho_i\omega_i \kappa_i L_i M_i}    \nonumber \\
&&=  \frac{1}{2}N_\omega \delta_{f,i} -\sqrt{\frac{3}{8}} \left (\CG{\omega_i \kappa_i L_i M_i}{(2\,0)00}{\omega_f \kappa_f L_f M_f} \RedME{\sigma n_f\rho_f\omega_f }{A^{(2\,0)}}{\sigma n_i\rho_i\omega_i} 
+{\rm conjugate} \right),
\end{eqnarray}
\end{widetext} 
where $\CG{\omega_i \kappa_i L_i M_i}{(2\,0)00}{\omega_f \kappa_f L_f M_f} $ is an \SU{3} Clebsch-Gordan coefficient. The matrix elements of the \SpR{3} generators, $A^{(2\,0)}$, $B^{(0\,2)}$, and $C^{(1\,1)}$, reduced with respect to \SU{3}, such as $\RedME{\sigma n_f\rho_f\omega_f }{A^{(2\,0)}}{\sigma n_i\rho_i\omega_i}$,  are known exactly \cite{RosensteelR83,RoweRC84,Rowe84,Hecht85,Rosensteel90} and the steps to calculate them are outlined in the appendix.

The simplest, \SpR{3}-preserving Hamiltonian, besides the  HO Hamiltonian ($H_0$), of importance to nuclear dynamics is \cite{Elliott58ElliottH62,BohrMottelson69},
\begin{equation}
H_{\rm E} = H_0-\frac{\chi}{2}Q.Q,
\end{equation}
with $\chi$ being a coupling constant and,
\begin{eqnarray}
&& H_0 = \sum_{i=1}^A \left(\frac{{\mathbf p}_{i}^2}{2m}+\frac{m\Omega^2 {\mathbf r}_{i}^2}{2}\right) \\
&& \half Q\cdot Q  =  \half \sum_{ij} q(i)\cdot q(j).
\end{eqnarray}
Here $q_{2M}(i) =\sqrt{16\pi/5b^4} r_i^2Y_{2M}(\hat {\mathbf r}_i)$ is the dimensionless  single-particle mass quadrupole moment and ${\mathbf r}_{i}$ (${\mathbf p}_{i}$) is the particle position (momentum) coordinate relative to the CM. 
In the limit of a single, valence shell ($N_n=0$), where \SU{3} becomes the relevant symmetry, the Hamiltonian $H_{\rm E}$ was shown to effectively describe rotational features of light nuclei in the framework of the established Elliott model \cite{Elliott58ElliottH62}. The success of such an effective nuclear interaction is not unexpected, as the spherical HO potential and the $Q.Q$ interaction directly follow from the second and third term, respectively, in the  long-range expansion of any two-body central force, e.g., like the Yukawa radial dependence,  
$V^{(2)}=\sum _{i<j} V(r_{ij}/a)=\sum _{i<j} (\xi_0+\xi_2 r^2_{ij}/a^2+\xi_2 r^4_{ij}/a^4+ \dots)$ \cite{Harvey68}, for a  range parameter $a$. However, in multi-shell studies, the attractive $Q.Q$ term  becomes ever stronger with increasing $N_n$ and starts to dominate the dynamics. Hence, $H_{\rm E}$ yields unphysical solutions.

A successful extension to  multiple shells has been achieved and applied to the $^{24}$Mg $g.st.$ rotational band \cite{PetersonH80}, where 
an interaction given as a polynomial in $Q$,  $Q\cdot Q$, $\left[ Q\times Q \right]\cdot Q$, and $(Q\cdot Q)^2$,  was employed. 

Furthermore, in multi-shell studies, the $Q.Q-\langle Q.Q\rangle_{N_n}$ interaction has been employed \cite{CastanosD89,RosensteelD85}, where $\langle Q.Q\rangle_{N_n}$ is the average contribution of $Q.Q$ within the subspace of $N_n$ HO  excitations, that is,  the trace of $Q.Q$  divided by the space dimension for a fixed $N_n$.  This removes the large monopole contribution of the  $Q.Q$ interaction, which, in turn, helps eliminate the considerable renormalization of the zero-point HO energy, while retaining the $Q.Q$-driven behavior of the wavefunctions. 

\subsection{No-core symplectic model with $H_{\gamma}$ for intermediate-mass nuclei}
We consider a novel effective many-nucleon interaction \cite{DreyfussLDDB13} suitable for large-$N_{\max}$ no-core shell models,
\begin{equation}
H_{\gamma}                     = H_0 + \frac{\chi}{2} \frac{\left( e^{-\gamma(Q.Q-\langle Q.Q\rangle_{N_n})} -1 \right)}{\gamma},
\label{effH}
\end{equation}
that addresses the limitations of the conventional $H_{\rm E}$,  while  retaining the $H_{\rm E}$ important features in the limit $\gamma \rightarrow 0$, 
where $\gamma$ is a positive adjustable parameter. We take $\chi=\ho/(4\sqrt{N_{\omega,f}N_{\omega,i}})$ with $N_{\omega, f(i)}$ the total HO quanta of the final (initial) many-body basis state. 
The decrease  of $\chi$ with $N_ \omega $, to a leading order in $\lambda/N_ \omega $, has been shown by Rowe \cite{Rowe67} based on self-consistent arguments and used in an \SpR{3}-based study of cluster-like states of $^{16}$O \cite{RoweTW06}. 
 
Above all, the effective interaction (\ref{effH}) introduces hierarchical many-body interactions in a prescribed way (for $\gamma \ll 1$). $H_\gamma$ also ensures that  the $Q.Q$ term  tails off for large $N_n$ eliminating its ever stronger attraction with increasing $N_n$. Such an interaction directly ties to the $Q$ polynomial considered in the above-mentioned study of Ref. \cite{PetersonH80}. Indeed, while higher-order terms in $Q\cdot Q$ of Eq. (\ref{effH}) could be understood as  a renormalization (as shown in Ref.  \cite{LeBlancCVR86}) of the $\chi$ coupling constant of the  $NN$ interaction, $-\frac{1}{2}\sum_{ij} q(i) \cdot q(j)$:
\begin{equation}
\frac{\chi}{2 \gamma}{\left( e^{-\gamma Q\cdot Q} -1 \right)}= -\frac{1}{2}[\chi(\sum_{k=0}^{\infty} \frac{(-\gamma )^{k}(Q \cdot Q)^{k}}{(k+1)!}) ]Q \cdot Q,
\end{equation}
 they become quickly negligible  for a reasonably small $\gamma$. E.g., we find that for $^{12}$C, besides $Q\cdot Q$, only one term is  sufficient for the ground-state band, while three terms are sufficient for the Hoyle-state band \cite{DreyfussLDDB13}.   
 \begin{figure}[t]
\begin{center}
(a) \hspace{1in} (b) \hspace{1in} (c) \\
\includegraphics[width=0.32\columnwidth]{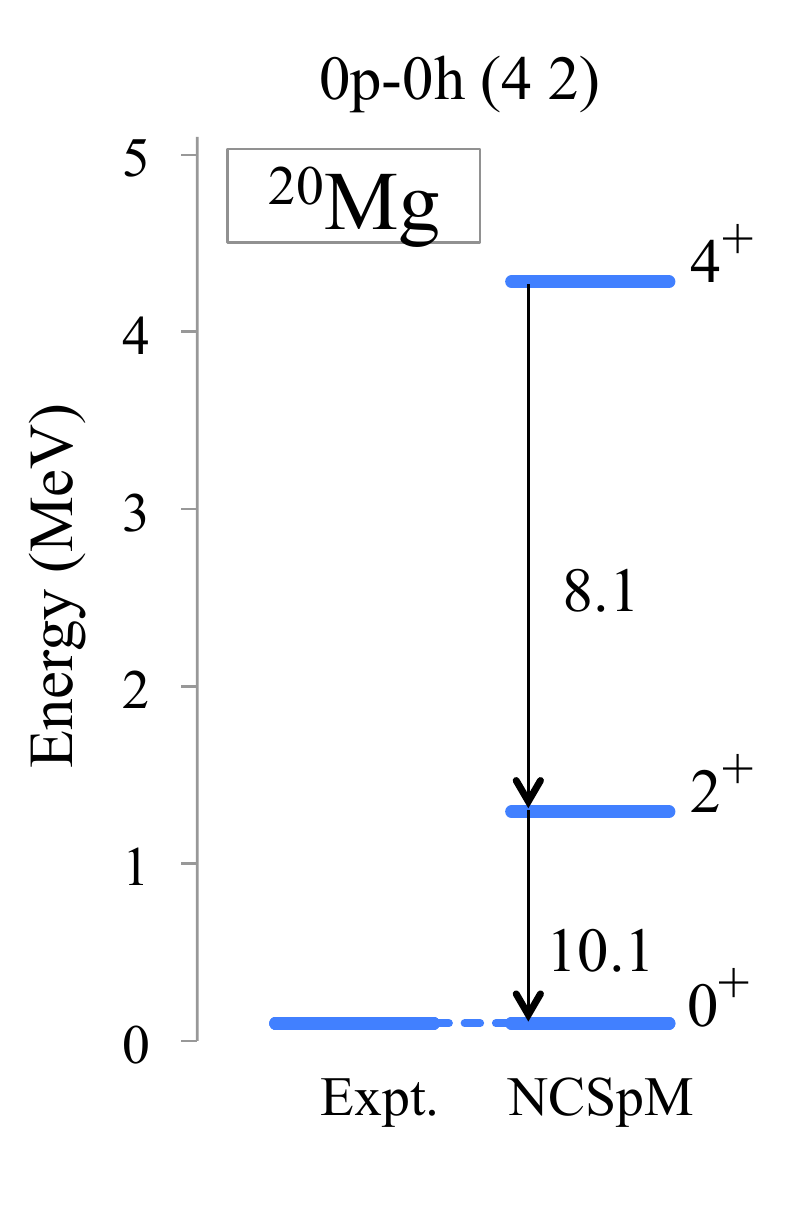}
\includegraphics[width=0.32\columnwidth]{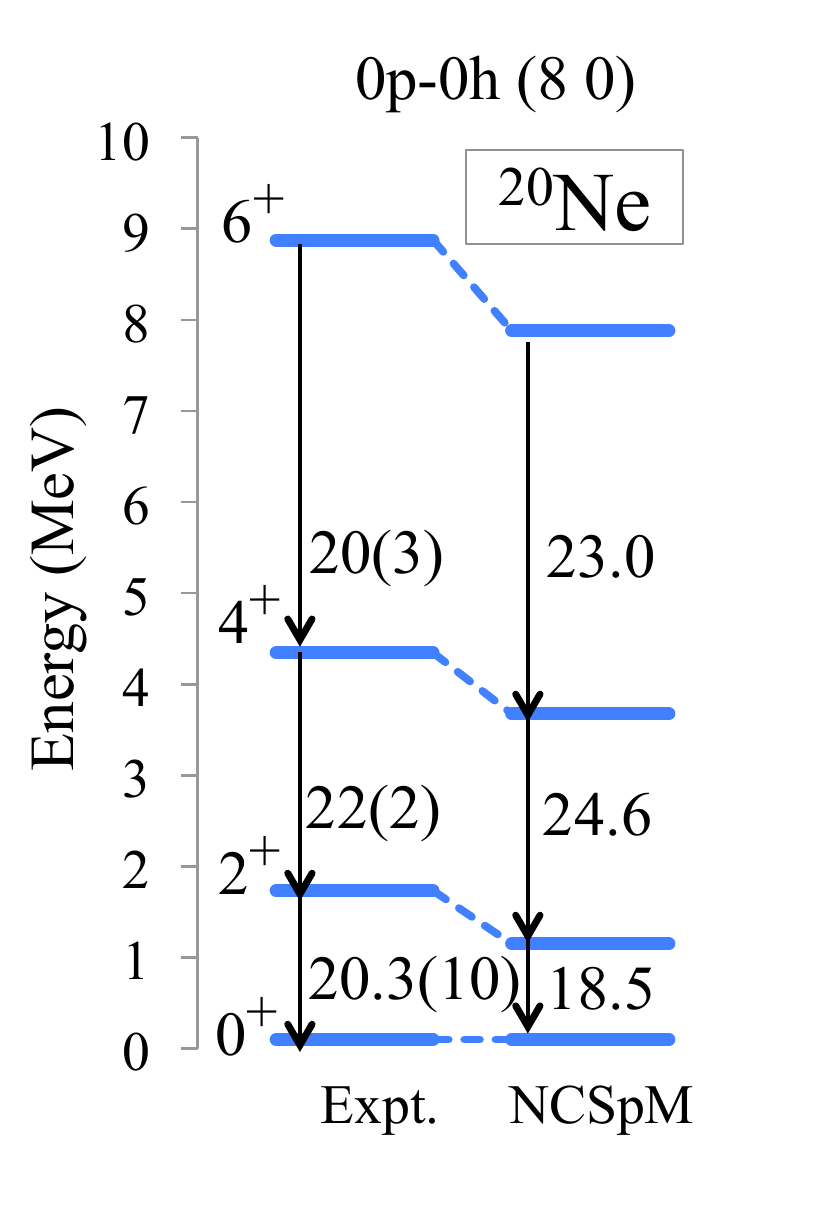}
\includegraphics[width=0.32\columnwidth]{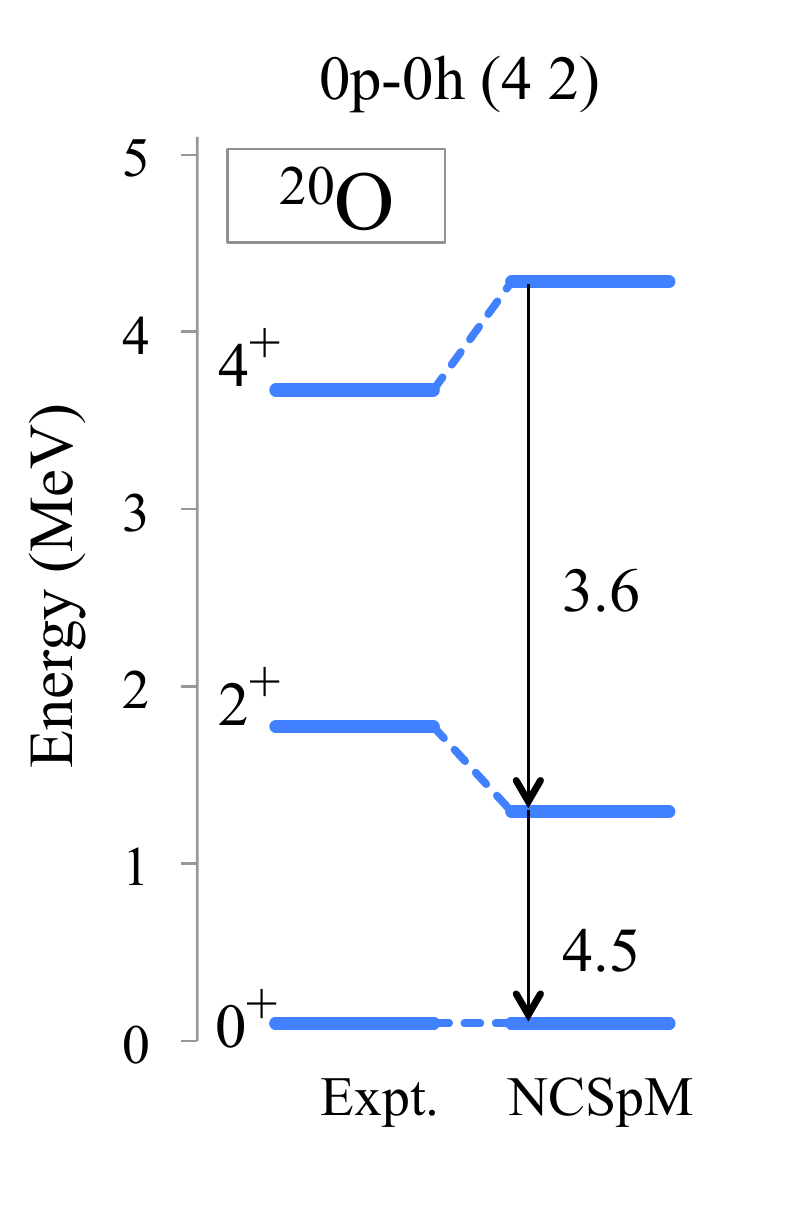}
\end{center}
\caption{
NCSpM energy spectrum of  (a) $^{20}$Mg  and (c) $^{20}$O using the $S_p=S_n=S=0$ $48.5 (4\, 2)$ \SpR{3} irrep built over the most deformed 0\ho~ bandhead, as well as of (b) $^{20}$Ne using the $S_p=S_n=S=0$ $48.5(8\, 0)$ \SpR{3} irrep  built over the most deformed 0\ho~ bandhead. Experimental data (``Expt.") is from \cite{Tilley98A20}. $B(E2)$  transition rates are in W.u. units.  
}
\label{enSpectrumNe20}
\end{figure}
\begin{figure}[t]
\includegraphics[width=0.45\textwidth]{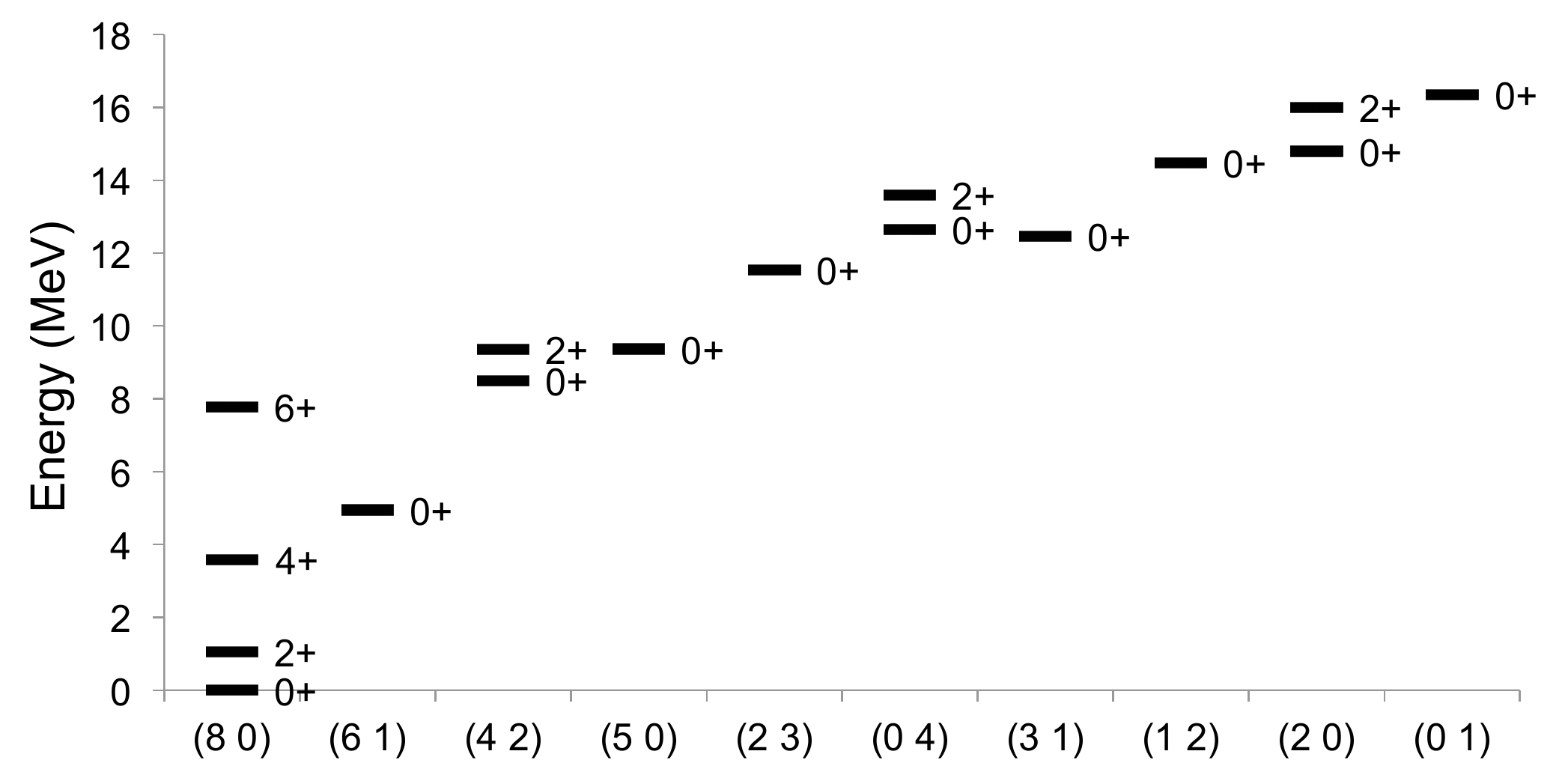}
\caption{
$^{20}$Ne low-lying states obtained by NCSpM for an $N_{\max}=12$ model space consisting of all possible 0\ho~ \ph{0} symplectic bandheads (for $S=0$ and $S=1$; spin-2 states are not shown). The lowest-lying $48.5(8\, 0)$ \SpR{3} irrep is selected for the $^{20}$Ne calculations (Fig. \ref{enSpectrumNe20}b).
}
\label{enSpectrumNe20full0hw}
\end{figure}
\begin{figure*}[th]
\begin{center}
(a) \hspace{2in} (b) \hspace{2in} (c) \\
\includegraphics[width=0.32\textwidth]{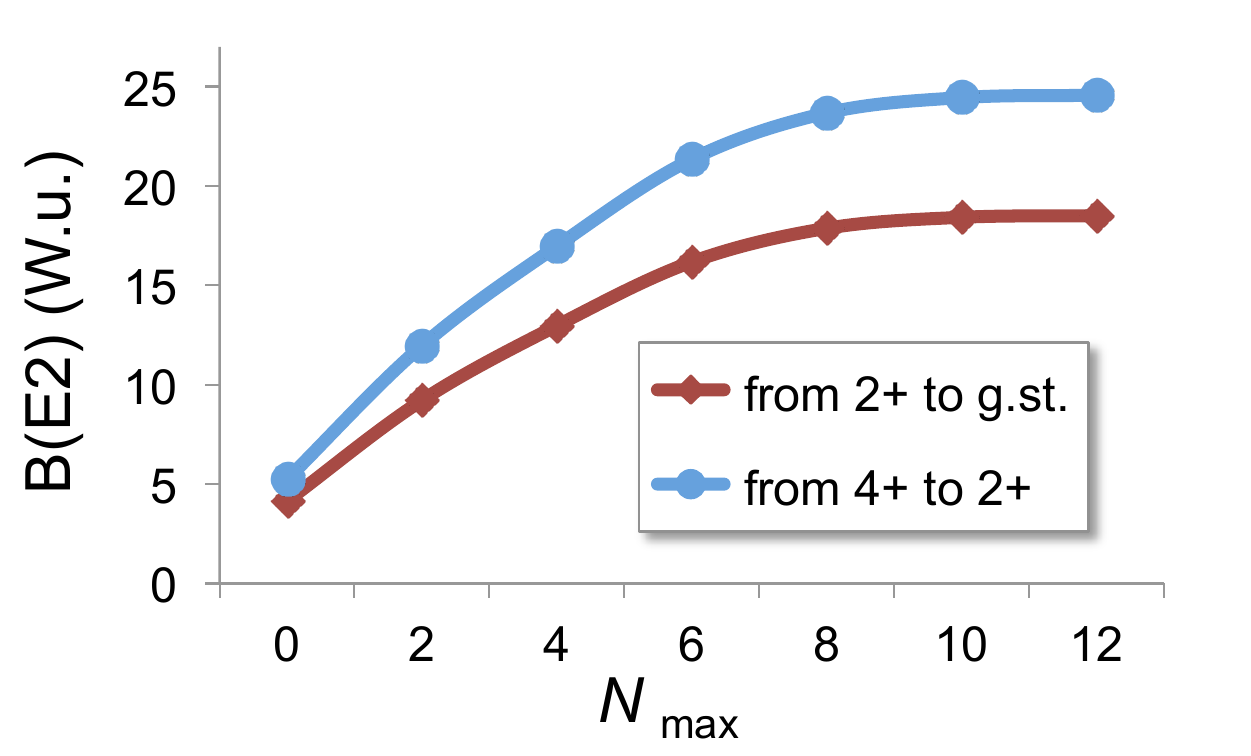}
\includegraphics[width=0.32 \textwidth]{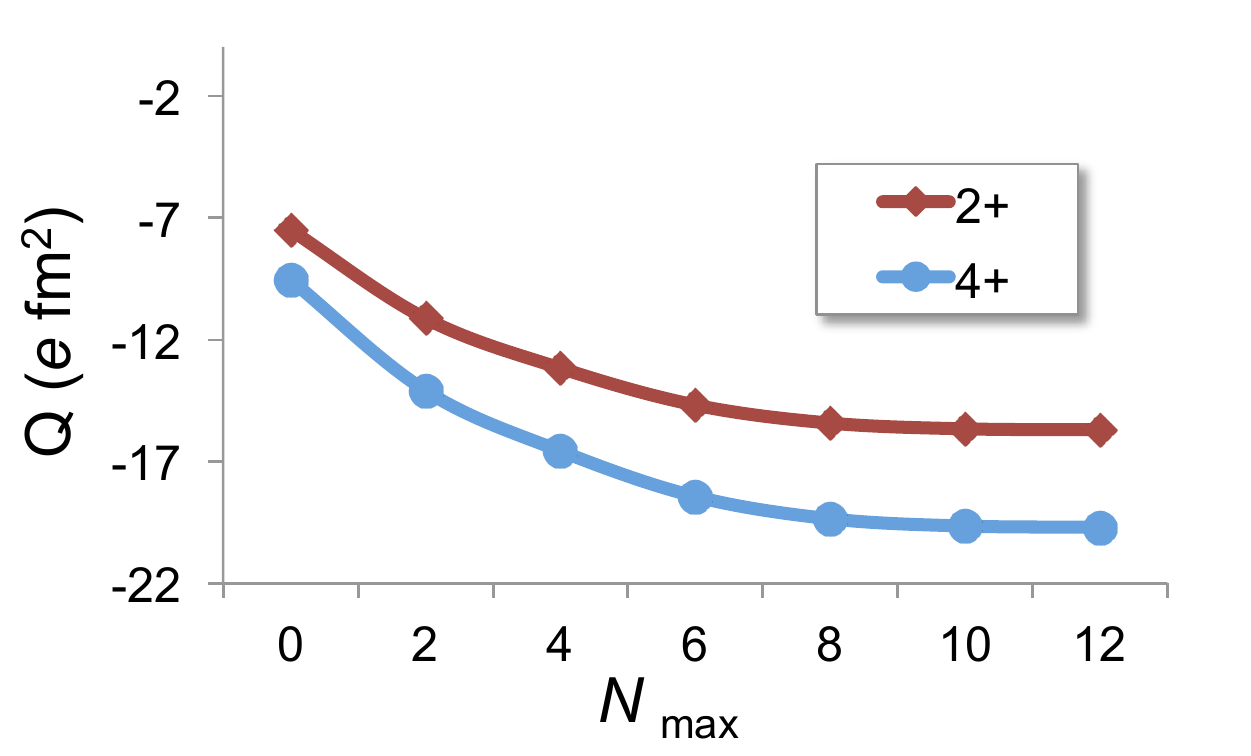}
\includegraphics[width=0.32 \textwidth]{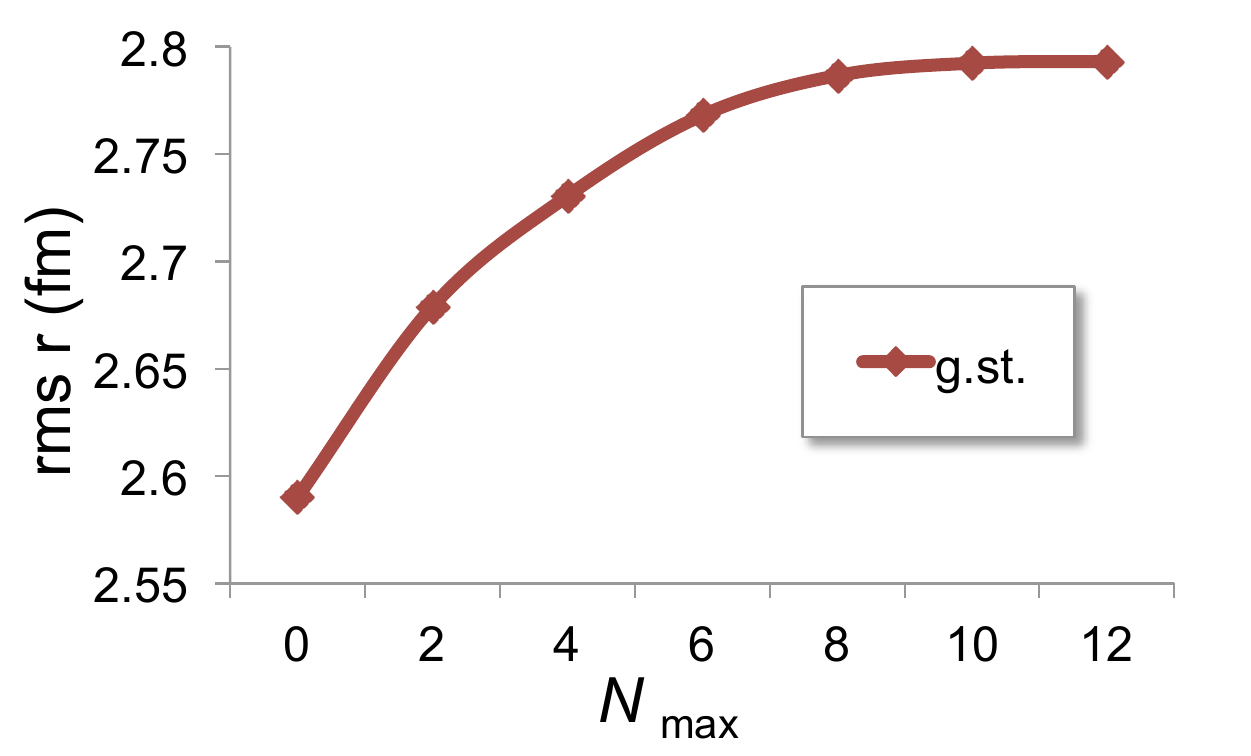}
\end{center}
\caption{
NCSpM observables for  $^{20}$Ne using the  \ph{0} $48.5(8\, 0)$  \SpR{3} irrep as a function of the model space, $N_{\max}$: (a) $B(E2;\, 2^+_1 \rightarrow 0^+_{g.st.})$  and $B(E2;\, 4^+_1 \rightarrow 2^+_1)$ transition strengths; (b) electric quadrupole moments for $2^+_1$ and $4^+_1$; and (c) the matter rms radius of the ground state.
}
\label{Ne20obsNmax}
\end{figure*}

For the NCSpM calculations, we use the empirical estimate $\ho \approx 41/A^{1/3}$, namely,  $\ho=18$ MeV for $A=12$ (for analysis of $^{12}$C \cite{DreyfussLDDB13}) and $\ho=15$ MeV for $A=20$ and $A=22$ isotopes. The \ho~ value, in turn, fixes the $\chi$ coupling strength of the $Q.Q$-term. With $\ho N_{\omega}$, the eigenvalue of $H_0$ in the model Hamiltonian (\ref{effH}), and with $\chi$ proportional to $\ho/4$, the eigenstates are rendered \ho-independent.

\section{Results and Discussions}
The NSCpM utilizes Bahri's symplectic computational code \cite{Bahri95_Sp3r_code} that uses Draayer \& Akiyama's numerical \SU{3} package \cite{AkiyamaD73}. The model has been successfully applied to the ground-state and Hoyle-state rotational bands in $^{12}$C \cite{DreyfussLDDB13}, where both rotational features and $\alpha$-cluster substructures have been described in the fully microscopic $N_{\rm max}=20$ no-core shell-model framework, as suggested by  the reasonably close agreement of  the model outcome with experiment and {\it ab initio} results in smaller spaces. The present study reveals that the model is also applicable to low-lying states of other light nuclei without any parameter adjustment, namely, we use $\gamma =0.74\times 10^{-4}$,  the value obtained in the NCSpM analysis for $^{12}$C. 

In particular, we focus on the $g.st.$ rotational band of selected $A=20, 22$ and $24$ isotopes. We note that, for the $g.st.$ band as opposed to cluster-driven excited rotational bands, comparatively lower $N_{\rm max}$ values are necessary to achieve convergence of energies, $E2$ observables and radii, with $N_{\rm max}=12$ found to be sufficient for the present calculations. 

Model spaces are down-selected based on findings of {\it ab initio} large-scale calculations for $^{12}$C and $^{16}$O  that have revealed low-spin and high-deformation dominance  \cite{DytrychDSBV09}, as well as the importance of symplectic irreps built over the most deformed 0\ho~ bandhead (the leading  \SU{3}  configuration) \cite{DytrychSBDV06}. For example, the latter study has shown a preponderance of the 0\ho~$(0\, 4)$ symplectic irrep  in $^{12}$C, which is indeed the irrep built over the  most deformed 0\ho~ bandhead, that is,  the spin-zero $(0\, 4)$.

\subsection{$^{20}$Ne and $A=20$ isotopes}
We present calculations for the $g.st.$ rotational band  of $^{20}$Ne together with the short-lived $^{20}$Mg at the proton-drip line (with no measured energy spectrum) and, its mirror nucleus, the neutron-rich $^{20}$O. They are indeed well described by the NCSpM in a $N_{\rm max}=12$  model space, where convergence of results is achieved (Fig. \ref{enSpectrumNe20} and Table \ref{Observables}). 

For $^{20}$Mg and $^{20}$O,  the model space is down-selected to only one spin-zero symplectic irrep, $(4\,2)$, for $J^\pi=0^+, 2^+$, and $4^+$, with 1299  basis states (fixed $M$).  For $^{20}$Ne, the model space consists of the spin-zero symplectic irrep, $(8\,0)$, for $J^\pi=0^+, 2^+, 4^+,$ and $6^+$, with 1070  basis states. All these irreps are built over the most deformed \ph{0} bandhead and expand up through $N_{\max}=12$.

To show the significance of the symmetry-based selection and the important role of  the  most deformed 0\ho~ bandhead  together with the \SpR{3} excitations thereof, we consider  a model space for $^{20}$Ne that consists of all symplectic irreps that start at 0\ho. The resulting NCSpM energy spectrum is displayed in Fig. \ref{enSpectrumNe20full0hw} for $S=0$ and $S=1$. Indeed, no other $0^+$ state is found to lie below the $0^+$ of $(8\,0)$  for the $\gamma$ parameter used here ($S=2$ symplectic bandheads render states higher than 8-9 MeV).  The $0^+$ of $(6\,1)$  is as much as 5 MeV above the $(8\,0)$, while all other $0^+$ states lie at $\sim 10$ MeV and higher. This indicates that the $(8\,0)$ irrep expanded up through $N_{\max}=12$ is indeed suitable for a reasonable description of the ground state of $^{20}$Ne.

An important feature of the NCSpM is that it provides electric observables without the need for introducing effective charges. And while $g.st.$ rotational energies converge comparatively quickly, at $N_{\max} \sim 4$, we find that larger model spaces are needed to reproduce observables sensitive to enhanced collectivity (Fig. \ref{Ne20obsNmax}). For $N_{\max} \sim 4$, observables, such as the $B(E2)$ transition strengths, electric quadrupole moments, and matter rms radii  have realized only  60\% of their total increase as compared to the $N_{\max}=0$ counterparts. Indeed, additional  $40-50\%$ are needed for the $B(E2)$ strengths and $20\%$ for the $Q$ moments to obtain converged values. To  reach convergence and to avoid the use of effective charges, at least $N_{\max} = 10$ is necessary, which is where results are also  found to compare reasonably to experiment (Fig. \ref{enSpectrumNe20} and Table \ref{Observables}). This suggests that the model successfully reproduces observables that are informative of the state structure and the long-range behavior of the wavefunctions. 
\begin{figure}[th]
(a)\\
\includegraphics[width=0.6 \columnwidth]{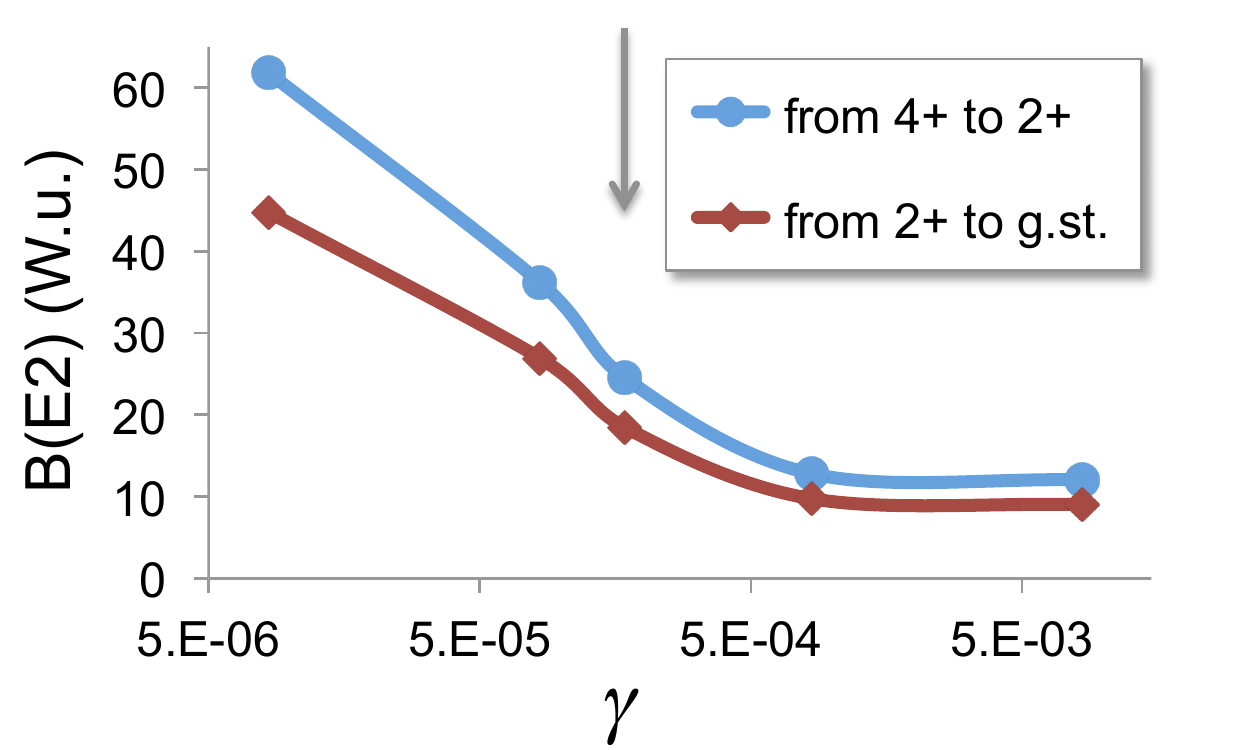} \\
(b) \\
\includegraphics[width=0.6 \columnwidth]{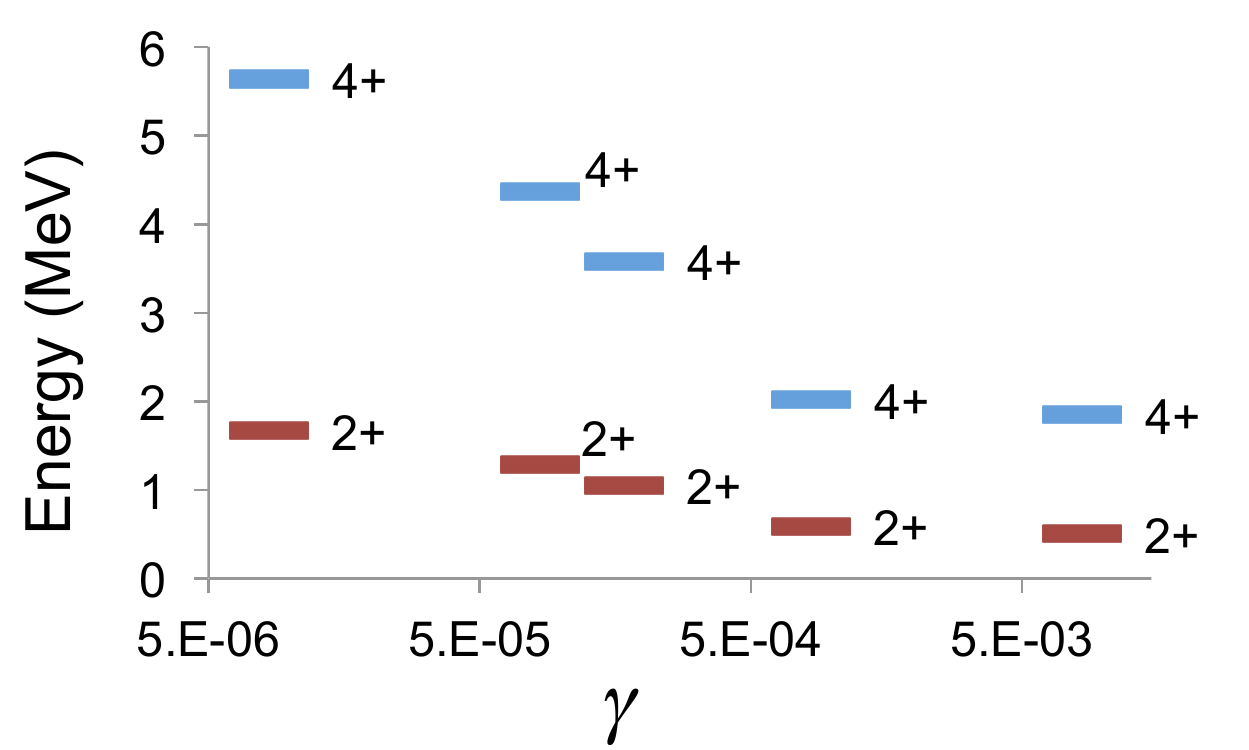} \\
\caption{
$\gamma$-Dependence of the $^{20}$Ne $g.st.$ rotational band: (a) $B(E2)$ transition strengths and (b) $2^+_1$ and $4^+_1$ energies. The grey vertical arrow indicates $\gamma=0.74\times 10^{-4}$, the value obtained in NCSpM analysis for $^{12}$C and used in the present calculations with no further adjustments.
}
\label{Ne20_gammaDependence}
\end{figure}

While the model parameter $\gamma$ has not been adjusted in the present study (but was significantly limited by the three lowest-lying $0^+$ states in $^{12}$C \cite{DreyfussLDDB13}), its value has a large effect upon observables under consideration.  A typical dependence on this parameter for $sd$-shell nuclei is shown for $^{20}$Ne (Fig. \ref{Ne20_gammaDependence}). As the $\gamma$ value decreases for a given nucleus, thereby increasing the tendency of the high-\ho~ excitations to become energetically favorable,  the nucleus expands spatially and the $g.st.$ rotational band  stretches energetically.  This is accompanied by enhancement of collectivity  and by considerably larger $B(E2)$ transition strengths. It is then remarkable that without adjusting $\gamma$ for $sd$-shell nuclei, energy spectra and other observables are found in a reasonable agreement with the experimental counterparts, as shown here for $A=20$ nuclei (Fig. \ref{enSpectrumNe20} and Table \ref{Observables}) and next for heavier isotopes.
\begin{table}[t]
\caption{NCSpM matter rms radii $r_m $ (fm) of the ground state and quadrupole moments $Q$ ($e$\,fm$^2$) of the $2^+$, $4^+$ and $6^+$ states of the $g.st.$ rotational band for the nuclei under consideration. Experimentally deduced matter radii  are summarized in Ref. \cite{OzawaST01} and each of the original references is provided in the table; measured $Q$ moments are taken from Refs. \cite{Tilley98A20,Firestone05A22} for $A=20$ and $22$, respectively.
}
\begin{center}
\begin{tabular}{llrrrr}
\hline\hline
&& \hspace{0.2in}$r_m (0^+_{\rm gs})$	& \hspace{0.4in}$Q_{2^+_1}$ & \hspace{0.3in}$Q_{4^+_1}$ & \hspace{0.3in}$Q_{6^+_1}$ \\
\hline
$^{20}$Mg	&	Expt.	&$	2.88(4)\tablenote{From Ref. \cite{OzawaST01ref28}} 	$&	--	&	--	&	--	\\
	&	NCSpM	&$	2.73	$&$	-12.67	$&$	-16.67	$&	--	\\
$^{20}$Ne	&	Expt.	&$	2.87(3)^a	$&$	 -23(3)	$&	--	&	--	\\
	&	NCSpM	&$	2.79	$&$	-15.69	$&$	-19.69	$&$	-21.05	$\\
$^{20}$O	&	Expt.	&$	2.69(3)^a	$& -- &	--	&	--	\\
	&	NCSpM	&$	2.73	$&$	-8.45	$&$	-11.11	$& --\\
$^{22}$Mg	&	Expt.	&$	2.89(6)\tablenote{From Ref. \cite{OzawaST01ref19}} 	$&	--	&	--	&	--	\\
	&	NCSpM	&$	2.82	$&$	-17.88	$&$	-23.07	$&$	-25.93	$\\
$^{22}$Ne	&	Expt.	&	--	&$	-17(3)	$&	--	&	--	\\
	&	NCSpM	&$	2.82	$&$	-14.90	$&$	-19.22	$&$	-21.61	$\\
$^{24}$Si	&	Expt.	&	--	&	--	&	--	&	--	\\
	&	NCSpM	&$	2.40	$&$	-14.38	$&$	-18.18	$&$	-19.75	$\\
$^{24}$Ne	&	Expt.	&$	2.79(13)\tablenote{From Ref. \cite{OzawaST01}} 	$&	--	&	--	&	--	\\
	&	NCSpM	&$	2.40	$&$	-10.27	$&$	-12.98	$&$	-14.11	$\\
\hline\hline
\end{tabular}
\end{center}
\label{Observables}
\end{table}%

\subsection{$A=22$ and $24$}
We perform NCSpM $N_{\max}=12$ calculations with no parameter adjustment (using $\ho=15$ MeV and $\gamma =0.74\times 10^{-4}$) for $^{22}$Mg and $^{22}$Ne. 
For these nuclei,  the $N_{\max}=12$ model space is down-selected to only one spin-zero symplectic irrep, $(8\,2)$, for $J^\pi=0^+, 2^+$, $4^+$ and $6^+$, with 2900  basis states (fixed $M$). 

\begin{figure}[th]
\includegraphics[width=0.44\columnwidth]{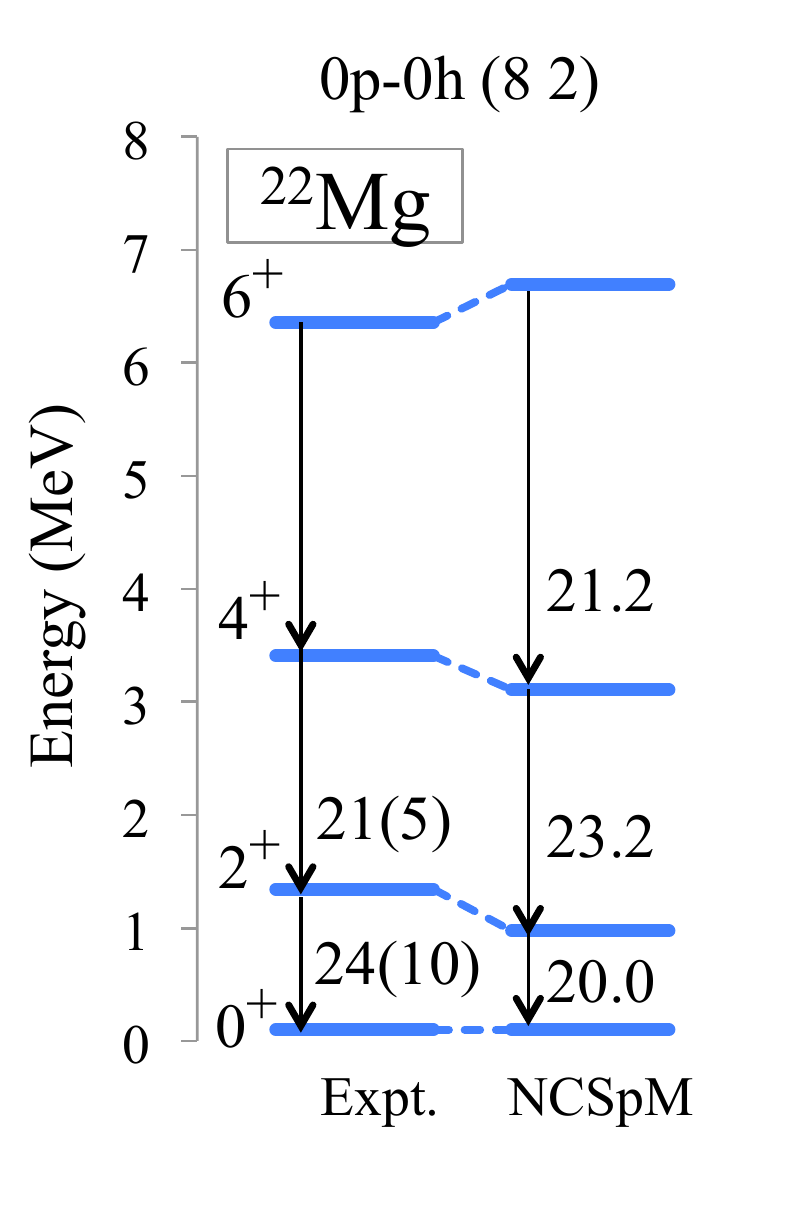}
\includegraphics[width=0.44 \columnwidth]{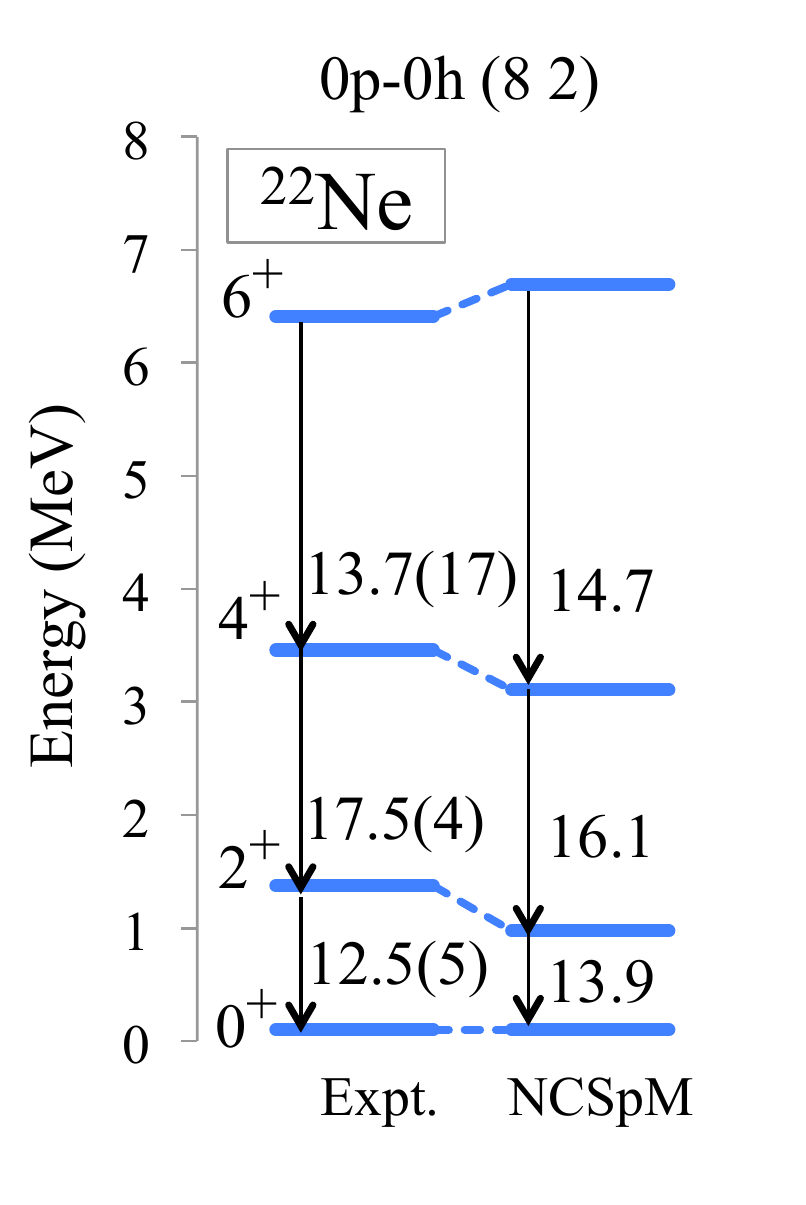}
\caption{
NCSpM energy spectrum of  $^{22}$Mg  and its mirror nucleus $^{22}$Ne  using the  $S_p=S_n=S=0$ $55.5(8\, 2)$ \SpR{3} irrep built over the most deformed 0\ho~ bandhead. Experimental data (``Expt.") is from \cite{Firestone05A22}.  $B(E2)$  transition rates are in W.u. units.
}
\label{enA22}
\end{figure}

Calculations for $^{22}$Mg and $^{22}$Ne  yield energy spectra in close agreement with experiment (Fig. \ref{enA22}). In addition, most of the $B(E2)$ transitions strengths (Fig. \ref{enA22}) as well as electric quadrupole moments and matter rms radii (Table \ref{Observables}) predicted  by the model  fall within the experimental uncertainties where measurements or experimentally deduced values exist.

\begin{figure}[th]
\includegraphics[width=0.44 \columnwidth]{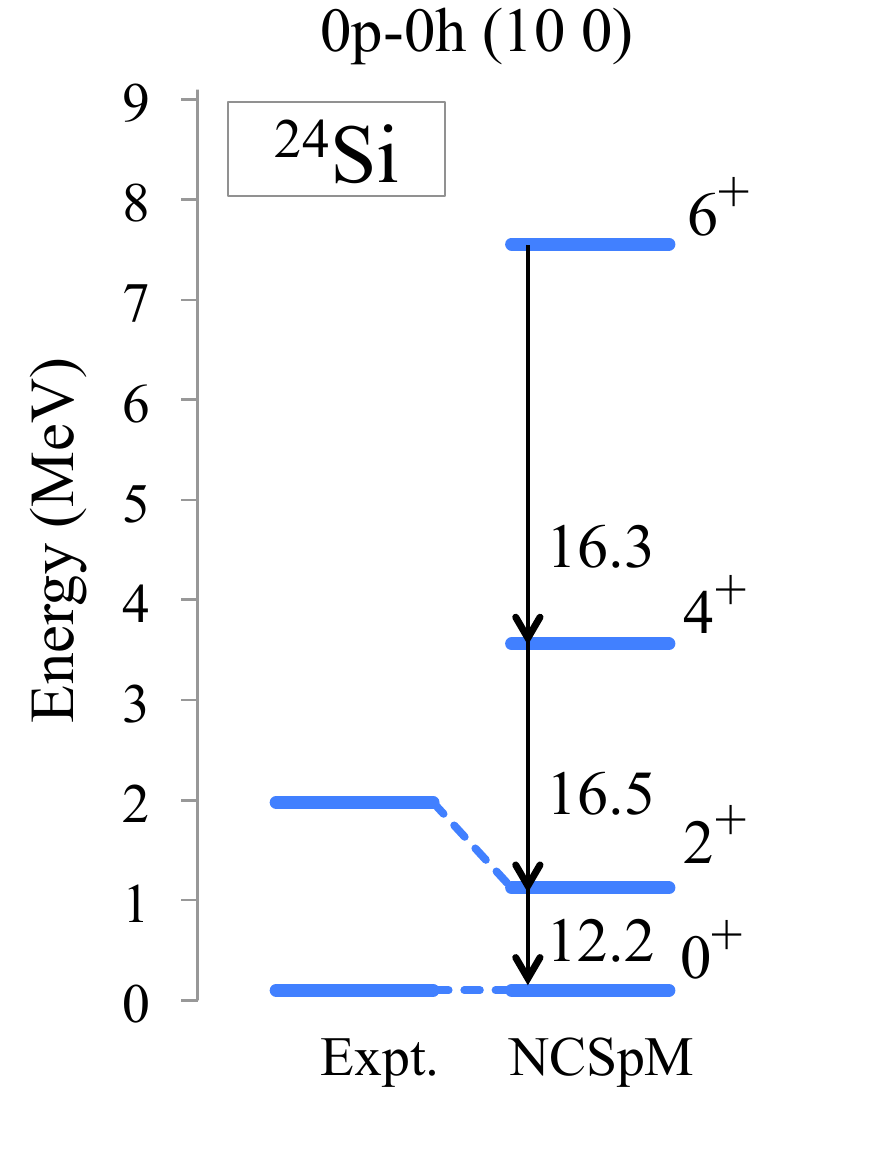}
\includegraphics[width=0.44\columnwidth]{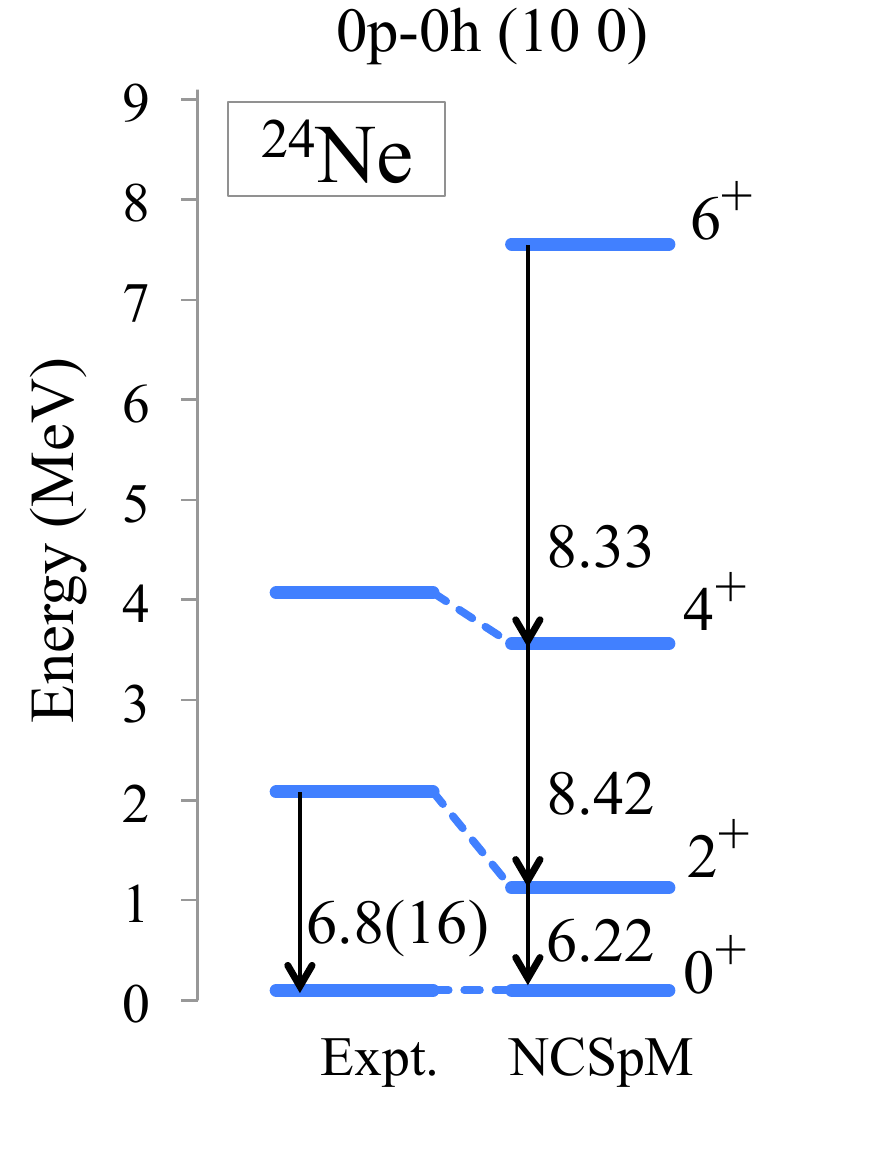}
\caption{
NCSpM energy spectrum of  $^{24}$Si and its mirror nucleus $^{24}$Ne using the $S_p=S_n=S=0$ $62.5(10\, 0)$ \SpR{3} irrep built over the most deformed 0\ho~ bandhead. Experimental data (``Expt.") is from \cite{Firestone07A24}.  $B(E2)$  transition rates are in W.u. units.
}
\label{enA24}
\end{figure}

\begin{figure}[th]
\includegraphics[width=\columnwidth]{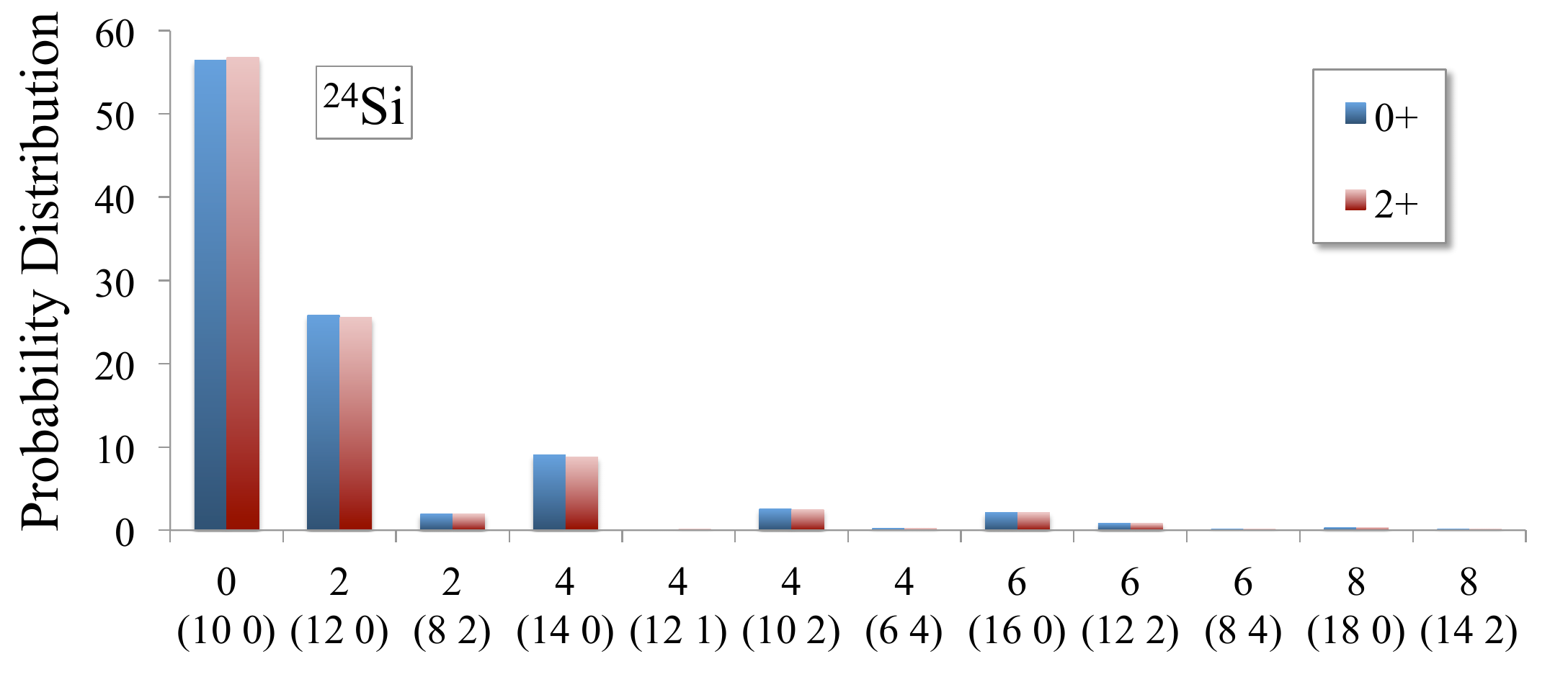}
\caption{
Probability distribution (in \%) for the $0^+$ ground state and the lowest $2^+$ state of  $^{24}$Si ($^{24}$Ne). Only contributions greater than 0.1\% are shown.
}
\label{Si24_probability}
\end{figure}

NCSpM $N_{\max}=12$ calculations (Fig. \ref{enA24}) are also presented here for the short-lived $^{24}$Si, even though a slightly larger $\ho$ value, 21 MeV,  is used only in this case. This nucleus is difficult to measure and knowledge on its structure, including spin-parity assignments and deformation of states, is necessary. For comparison, we also study the mirror nucleus $^{24}$Ne, for which richer experimental data is available.
For these nuclei,   the model space is down-selected to only one spin-zero symplectic irrep, $(10\,0)$, for $J^\pi=0^+, 2^+$, $4^+$ and $6^+$, with 1171  basis states (fixed $M$). The results, including energy spectra, $E2$ observables and radii, are found reasonable as compared to the available experiment (Fig. \ref{enA24} and Table \ref{Observables}). The $^{24}$Si wavefunctions, calculated by NCSpM  and independent of the choice for $\ho$, are found to be dominated by the $(10\,0)$ 0\ho-,  $(12\,0)$ 2\ho- and $(14\,0)$ 4\ho-configurations (Fig. \ref{Si24_probability}), thereby, as discussed in the following section, carrying considerably large prolate deformation. While there are 96 (or 274) basis states in the $(10\,0)$ symplectic irrep for the $0^+$ (or $2^+$) state, only a few of them contribute to the wavefunction at a level greater than 0.1\%, as shown in Fig. \ref{Si24_probability}. We note that the slightly smaller radius calculated by the model for  $^{24}$Ne suggests that additional spin-zero and spin-one irreps besides the $(10\,0)$ vertical cone are likely to  influence the low-energy dynamics. However, they are expected to remain of secondary importance to $(10\,0)$.

\section{Dominant deformed configuration}
Examinations of the \SU{3} content of the NCSpM wavefunctions bring forward important information  on deformation and associated dominant configurations through the deformation-related $(\lambda_\omega \,\mu_ \omega)$.
This is based on the mapping \cite{CastanosDL88} between the shell-model
$(\lambda \,\mu)$ \SU{3} labels  (microscopic) and the shape variables of the
Bohr-Mottelson collective model \cite{BohrMottelson69}, which provides a description of the nuclear surface in terms of the elongation
$\beta>0$ and the  $0 \le \gamma \le \pi/2$ asymmetry parameter. 
Specifically, in the limit of large deformation, $(\lambda\ 0)$ and $(0\ \mu)$  can be associated with a prolate
 ($\gamma = 0 ^\circ $) and  oblate  ($\gamma = 60 ^\circ$) shapes,
respectively, while larger $\lambda$ ($\mu$) values are linked to larger deformation, $\beta$. 
\begin{figure}[th]
\includegraphics[width=0.64\columnwidth]{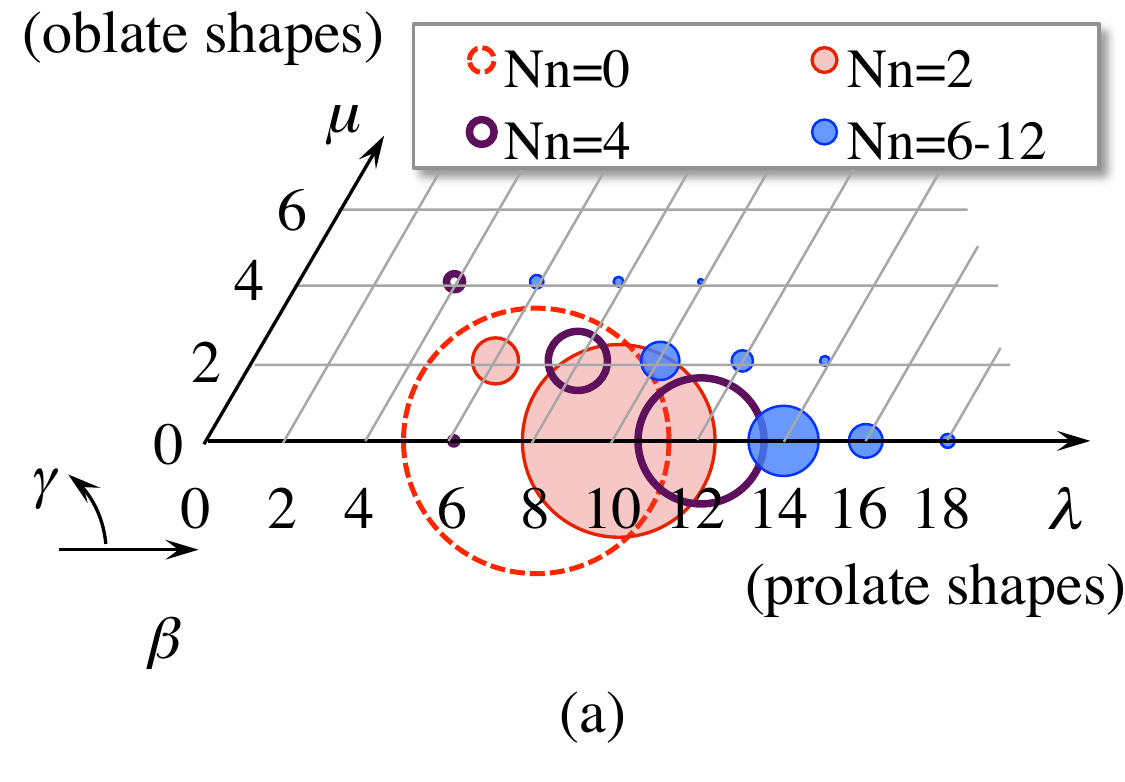}
\includegraphics[height=1.5 in]{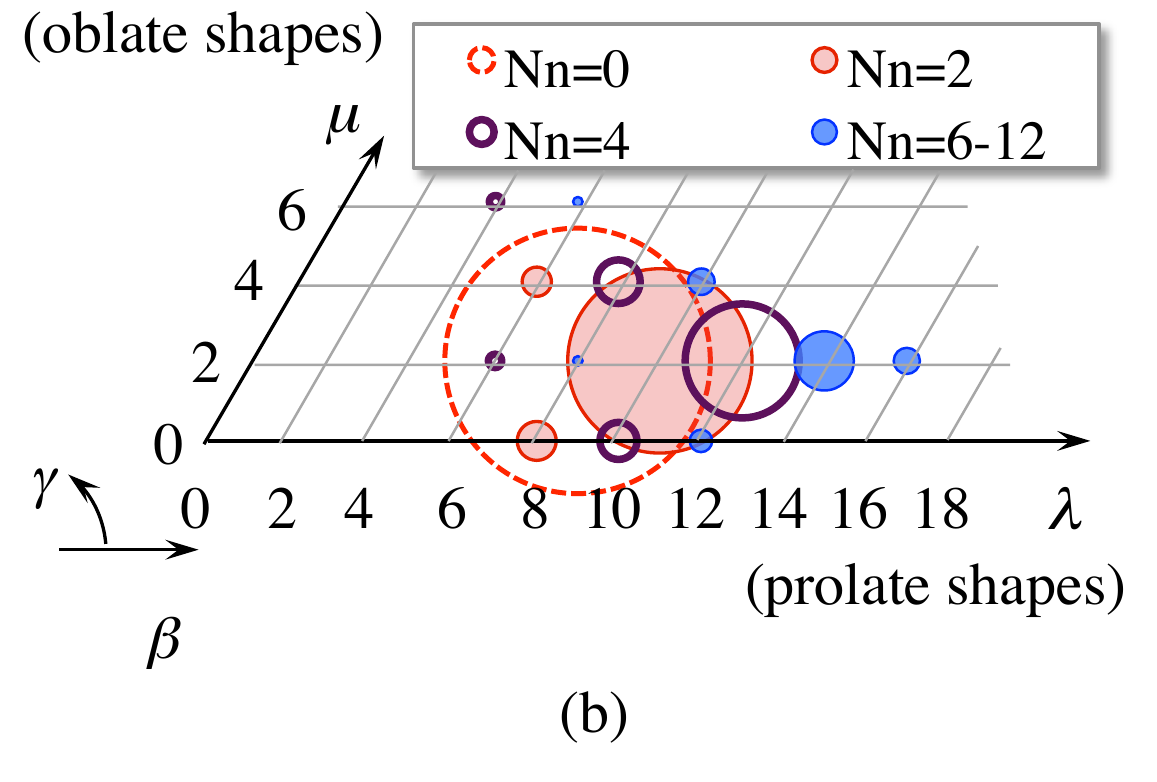}
\caption{
NCSpM probability distribution (specified by the area of the circles) for the ground state of  (a) $^{20}$Ne using the $(8\,0)$ \SpR{3} irrep, and (b) $^{22}$Ne using the $(8\,2)$ \SpR{3} irrep.
The symplectic states are grouped according to their $(\lambda_\omega \,\mu_\omega)$ \SU{3} symmetry, which are mapped onto the  $(\beta \,\gamma)$ shape variables  of the collective model (see text for further details). 
}
\label{probability}
\end{figure}

It is then clear that the nuclei under consideration are highly deformed and prolate in their low-lying states, as manifested in Fig. \ref{probability} for selected nuclei (this feature is the least pronounced for $^{20}$O and $^{20}$Mg). This is also confirmed by the negative large electric quadruple moments (Table \ref{Observables}) and enhanced $B(E2)$ values.  Fig. \ref{probability} further reveals that the most dominant modes are observed among the ones with high  $\lambda_\omega$ and low $\mu_ \omega$. Configurations in higher-\ho~ model spaces tend to increase the deformation (larger $\beta$) and decrease nuclear triaxiality (smaller $\gamma$) as compared to the predominant 0\ho~ configuration (the bandhead). These high-\ho~ configurations ($N_n =6$ and beyond), while contributing only slightly to the wavefunctions, bring in substantially large deformation, thereby becoming critical for the convergence  of the observables shown in Fig. \ref{Ne20obsNmax}.  

\section{Conclusion}
We carried forward a  no-core symplectic  NCSpM  study with a schematic long-range many-nucleon interaction that showed how highly deformed structures in intermediate-mass nuclei emerged out of a no-core shell-model framework. 
While previously the NCSpM has been successfully employed in a study of the  $\alpha$-cluster driven Hoyle-state rotational band in $^{12}$C  \cite{DreyfussLDDB13}, which has fixed the only adjustable parameter in the schematic interaction, here we show that the framework is extensible to low-lying states of other nuclei without any parameter adjustment. We focused on the $g.st.$ rotational band of $^{20}$Ne (a nucleus multiple of an $\alpha$ particle), as well as  of $^{22,24}$Ne, $^{20}$O and of the proton-rich $^{20,22}$Mg, and $^{24}$Si nuclei.  

By utilizing the symplectic symmetry, we were able to accommodate model spaces up through 15 major HO shells and hence, to take into account essential high-\ho~particle excitations.  These excitations were found key to the description of large deformation and the convergence of electric observables without effective charges.  These configurations were included in the shell-model space by considering only one symplectic irrep (vertical cone) built on the most deformed spin-zero bandhead and extended to $N_{\rm max}=12$. This further confirms the dominance of low-spin/high-deformation and the importance of the symplectic symmetry to the low-energy nuclear dynamics.

Most importantly, the NCSpM has allowed us to identify,  from a no-core shell-model perspective, components of the inter-nucleon interaction and type of particle excitations that appear foremost responsible for unveiling the primary physics governing highly-deformed structures,  starting with rotational and alpha-clustering features in the case of $^{12}$C \cite{DreyfussLDDB13} and $^{8}$Be \cite{LauneyDDTFLDMVB12}, but also expanding to the region of the lower $sd$ shell ($A\le24$). Therefore, the NCSpM appears as a useful tool to inform properties of the inter-nucleon interaction and to suggest efficacious shell-model truncation strategies to be employed in  {\it ab initio} studies.
\newline

\acknowledgements
We thank George Rosensteel and David Rowe  for useful discussions.
This work was supported by the U.S. NSF (OCI-0904874), the U.S. DOE (DE-SC0005248), and SURA, and in part by U.S. DOE (DE-FG02-95ER-40934).
ACD  acknowledges  support by the U.S. NSF (grant 1004822) through the REU Site in Dept. of Physics \& Astronomy at LSU.
\newline

\appendix*{{\bf Appendix} }

The \SU{3}-reduced matrix elements of the \SpR{3} generators are analytically known  \cite{RosensteelR83,Rowe84,Hecht85,Rosensteel90}. The steps to compute $\RedME{\sigma n_f\rho_f\omega_f }{A^{(2\,0)}}{\sigma n_i\rho_i\omega_i}$, similarly for $B^{(0\,2)}_{\mathcal{L}M}=(-)^{\mathcal{L}-M}(A^{(2\,0)}_{\mathcal{L}-M})^{\dagger}$, are outlined in what follows:
\begin{enumerate}
\item Calculations of non-normalized $\RedMEnonNorm{n_f}{ {\mathcal A}^{(2\,0)}}{ n_i}$ using Eq. (4.51) of Ref. \cite{Rowe85} with $n_1= \frac{N_n + 2 \lambda_n + \mu_n}{3} $, 
$n_2 = \frac{N_n - \lambda_n + \mu_n}{3} $, and $n_3 = \frac{N_n - \lambda_n -2\mu_n}{3}$ associated with $n_{i}=N_{n,i}\left(\lambda_{n,i}\,\mu_{n,i}\right)$ and $n_{f}$,
together with the notation, ${\mathcal A}^{(2\,0)}\rightarrow a^\dagger$;

\item Calculations of non-normalized $\RedMEnonNorm{\sigma n_f\rho_f\omega_f }{A^{(2\,0)}}{\sigma n_i\rho_i\omega_i}$ from   $\RedMEnonNorm{n_f}{{\mathcal A}^{(2\,0)}}{n_i}$ using Eq. (4.50) of Ref. \cite{Rowe85};

\item Calculations of  $\RedME{\sigma n_f\rho_f\omega_f }{A^{(2\,0)}}{\sigma n_i\rho_i\omega_i}$ from the non-normalized reduced matrix elements (step 2) using  the ${\mathcal  K}$-matrix approach \cite{Rowe84,Hecht85}. The present calculations utilize the full ${\mathcal  K}$ matrix (exact calculations). However, in the multiplicity-free case ($\rho_i^{\max}=\rho_f^{\max}=1$) or in the limit of large $\sigma$ \cite{RoweRC84}, the normalization matrix reduces to normalization coefficients (a diagonal ${\mathcal K}$ matrix) given by Eq. (17) of Ref. \cite{RoweRC84}.  
\end{enumerate}

For the $C^{(1\,1)}_{\mathfrak{L}M}$ \SU{3}-reduced matrix elements, see, e.g., Eq. (19) of Ref.  \cite{Rosensteel90}.  Using the reduced matrix elements of the \SpR{3} generators and the relation (\ref{Qgen}), the analytical formula for the $Q\cdot Q$ matrix elements has been derived in Ref. \cite{Rosensteel90}.

\end{document}